%% file: main.tex
\documentclass[conference]{IEEEtran}
\IEEEoverridecommandlockouts
\usepackage{amsmath,amssymb,amsfonts}
\usepackage{algorithmic}
\usepackage{graphicx}
\usepackage{textcomp}
\usepackage{siunitx}
\usepackage[capitalise, noabbrev]{cleveref}
\crefformat{equation}{(#1)}
\usepackage{glossaries}
\usepackage{booktabs}
\usepackage{multirow}
\usepackage{bm}			% for \bm ... bold math 
\usepackage{flushend}   % if the balance package fails

\usepackage{amsmath}

\def\innervector(#1,#2,#3){\ensuremath{\left( #1,#2,#3 \right)}}

\usepackage[backend=biber,style=ieee,maxnames=2, minnames=1]{biblatex}

\AtBeginBibliography{\footnotesize}

\loadglsentries{abbr}

\usepackage[dvipsnames,table,xcdraw]{xcolor}
\def\BibTeX{{\rm B\kern-.05em{\sc i\kern-.025em b}\kern-.08em
    T\kern-.1667em\lower.7ex\hbox{E}\kern-.125emX}}
\begin{document}\sloppy

\newcommand{\liesbet}[1]{{\color{ForestGreen}[Liesbet: #1]}}
\newcommand{\lieven}[1]{{\color{Green}[Lieven: #1]}}
\newcommand{\bert}[1]{{\color{RedViolet}[Bert: #1]}}
\newcommand{\jarne}[1]{{\color{Blue}[Jarne: #1]}}
\newcommand{\ben}[1]{{\color{PineGreen}[Ben: #1]}}
\newcommand{\gilles}[1]{{\color{LimeGreen}[Gilles: #1]}}

\DeclareSIUnit{\dBm}{dBm}	                % SI unit "dBm"
\DeclareSIUnit{\dBi}{dBi}                   % SI unit "dBi"
\DeclareSIUnit{\dBsm}{dBsm}                 % SI unit "dBsm"

% Math:
\newcommand{\herm}{\mathsf{H}}
\newcommand{\trp}{\mathsf{T}}

\title{Keeping Energy-Neutral Devices Operational:\\ a Coherent Massive Beamforming Approach\\
\thanks{Reported results from the REINDEER project,  funded by the European Union's Horizon 2020 RIA programme grant agreement No. 101013425.}
}

\vspace{-12pt}

% \and
% \and
% \and
% \and
% \and

\author{
    \IEEEauthorblockN{Jarne Van Mulders\IEEEauthorrefmark{1}, Bert Cox\IEEEauthorrefmark{1}, Benjamin J.\,B. Deutschmann\IEEEauthorrefmark{2}, Gilles Callebaut\IEEEauthorrefmark{1},\\ Lieven de Strycker\IEEEauthorrefmark{1} and Liesbet Van der Perre\IEEEauthorrefmark{1}}
    \IEEEauthorblockA{\IEEEauthorrefmark{1}\textit{KU Leuven,} Belgium, jarne.vanmulders@kuleuven.be, \IEEEauthorrefmark{2}\textit{Graz University of Technology,} Austria}
}
\maketitle
\begin{abstract}

Keeping the batteries on the shelf: this is the holy grail for low-cost \gls{iot} nodes. In this paper we study the potential of \gls{rf}-based wireless power transfer implementing coherent beamforming with many antennas to realize this ambitious target. We optimize the deployment of the antennas to charge \glspl{esl}, considering actual regulatory constraints. The results confirm the feasibility to create power spots that are sufficient to keep the high density of battery-less devices operational.

%This document is a model and instructions for \LaTeX.This and the IEEEtran.cls file define the components of your paper [title, text, heads, etc.]. *CRITICAL: Do Not Use Symbols, Special Characters, Footnotes, or Math in Paper Title or Abstract.
\end{abstract}


% \glsresetall

\section{Introduction}
%\liesbet{Liesbet to draft, probeer 1 kolom}
\Gls{en} devices can be defined~\cite{fi14050156} as active or passive devices for which the energy they are able to harvest from their environment ($E_{in}$) is at least as large as the energy they need for their consumption operations ($E_{cons}$). They are of great interest for many applications as they can theoretically remain operational forever without batteries or power plug.  
\Gls{rf}-based \gls{wpt} presents an attractive offer to remotely charge these devices~\cite{Mag2015}. However, the efficiency of conventional systems is very low, limiting practical applications within actual regulatory RF constraints.\newline
Distributed large antenna systems realizing beam focusing have been proposed to increase the efficiency of \gls{rf}-based power transfer drastically~\cite{Choi_distr}. 
In particular, physically large or distributed antenna infrastructures hold great potential for efficient yet regulatory-compliant power transfer that is achieved through operation in the array near field~\cite{Lopez2022,Deutschmann23ICC, Zhang23NF}.

Still, the question remains whether the efficiency improvements can suffice to power \gls{en} devices in realistic use cases where the regulatory constraints need to be adhered to, and potentially a high density of devices is present. In this paper, we study the feasibility of the approach for a representative case of \acrfullpl{esl}. As it is evident from basic link budget considerations that the antennas have a crucial impact on the received power, we study how many antennas are required, and where and how to deploy them considering their radiation pattern. In what follows, we assume that general prerequisites such as the initial access of \glspl{esl} and phase coherence between different transmitters are fulfilled. \newline
This paper is organized as follows. In the next section, we present the system set-up and methodology. The results are provided in Section~\ref{sec:results}. The conclusions are summarized and directions for future work are suggested in Section~\ref{sec:conclusions}.

% Belangrijk element: wat doen we met de antennes
% \begin{itemize}
%     \item Plaatsing
%     \item Aantal
%     \item Stralingspatroon
% \end{itemize}

\begin{figure}[ht]
    \centering
    \includegraphics[width=0.4\textwidth]{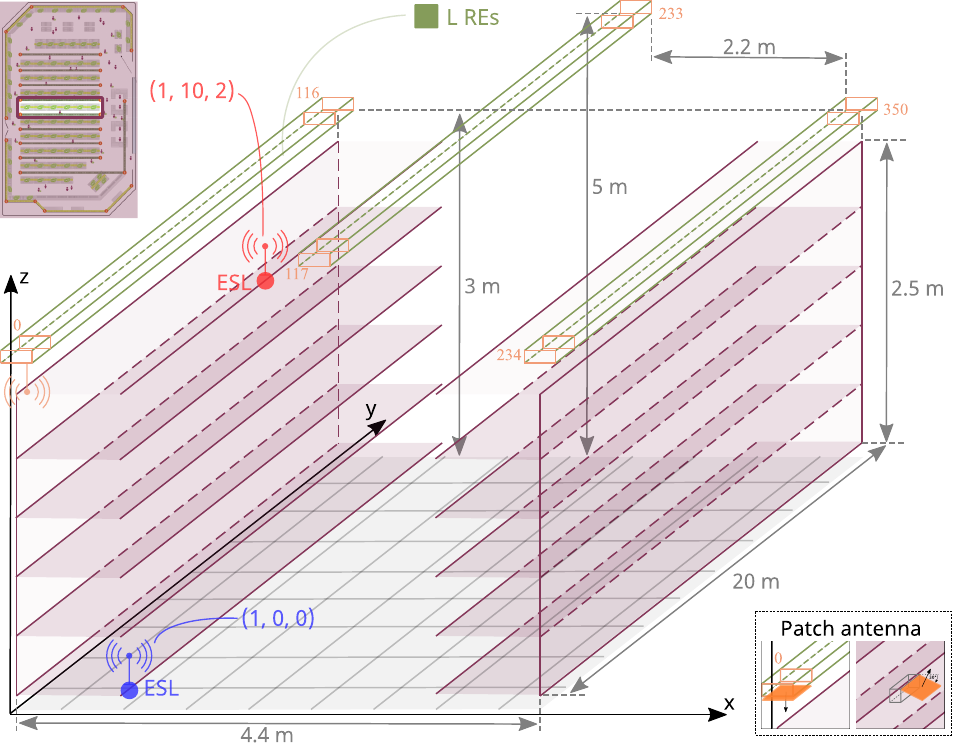}
    \caption{3D representation of a supermarket aisle with 351 antennas transmitting on \SI{868}{\mega\hertz}. The orange digits represent the antenna numbering.}
    \label{fig:aislestore}
\end{figure}

% test \cite{RemoteIoT_art}

\section{System design and methodology}
Coherent massive beamforming holds a potential for spectacular efficient improvements in wireless communication and power transfer~\cite{Kashyap2015}. Key parameters in a \gls{wpt} system are the necessary power density (\si{\milli \watt / \meter \squared}), the device density (per \si{\meter \squared}), the required DC-power at the receiver (\si{\milli \watt}) and the carrier frequency of the wireless link (\si{\giga \hertz}). A representative challenging use case scenario is found in retail, e.g., in supermarkets, where \glspl{esl} could be wirelessly powered. The goal of this research is to define and estimate the required transmit power radiated by antennas to power the \glspl{esl}, so they can perform a screen update twice a day. Additionally, different RF energy provisioning methods are compared to test the hypothesis that coherent massive beamforming is the most efficient way to keep the batteries on the shelf. %\lieven{how much energy xx J per update ?, also see my remark about mentioning my numbers already here} 

\subsection{Deployment Scenario}% and Energy Requirements}
\label{subsection:deploy}
%Different deployment scenarios such as tennis arena, production hall, supermarket, large apartment and patient care were introduced in D1.1. It would take us too far to comprehensively simulate and discuss all use cases. For simplicity, we further in this report consider the most challenging deployment scenario based on the assessment presented in \cref{tab:usecases}, namely the supermarket equipped with \glspl{esl} is considered as a representative. T
In this paper we consider a supermarket use case with %\jarne{REDUNDANT: a device density of \SI{20}{ESLs/\meter^2} and} 
up to 600 \glspl{esl} in an aisle. To clarify the study, we focus on a single aisle with three antenna arrays that can power the \gls{en} devices. Fig.\,\ref{fig:aislestore} shows a 3D sketch of the shop aisle with dimensions \SI{20}{\meter} by \SI{4.4}{\meter} and \SI{2.5}{\meter} high cabinets with %6, 
\SI{1}{\meter} wide shelves. In the depicted aisle, the three antenna arrays, with M $\lambda$/2 spaced antennas, are deployed above the cabinets and shown in green. Using an \SI{868}{\mega\hertz} carrier frequency, $M = 117$ antennas in each array can be placed in the \SI{20}{\meter}-long aisle. In total, we consider $L = 3M = 351$ available antennas in this particular case for one aisle. %The number of transmit antennas is indicated as $L$ ($= 3M$), with M the number of antennas in each array.
Together with \gls{esl} energy requirements described in \cref{subsection:requirements}, an estimate of the transmit power of each antenna can be determined. We consider limitations in radiated power by complying with the harmonised European standard concerning the \gls{rfid} equipment~\cite{ETSI_EN_302_208}. %\jarne{standaard citeren?}\liesbet{zou ik doen. is het een standaard of een regulation?}. 
%The total radiated power of all energy transmitters is required to be lower than the maximum allowable transmit level as prescribed by the standard.

\subsection{Energy Requirements}
\label{subsection:requirements}

The daily required DC energy is related to the \gls{esl} refresh rate and the energy to perform an \gls{esl} screen update. The largest energy consumer in an \gls{esl} device is the E-Ink Raw Display. In \cite{epaperscreen}, it was stated that it takes \SI{450}{\milli\joule} (\SI{15}{\second} $\times$ \SI{30}{\milli\watt}) to update the display. Additional energy ($\sim$\SI{50}{\milli\joule}) is needed to power the \gls{mcu} and demodulate the downlink information signal. In total, \SI{500}{\milli\joule} is required for one \gls{esl} screen update. Assuming this happens twice a day, approximately \SI{1}{\watt\second/day/\gls{esl}} of energy should be delivered. This corresponds to a constant net power of \SI{12}{\micro\watt}. Typically, \glspl{esl} are installed at a density of \SI{20}{ESLs\, /\meter^2}. Consequently, they together necessitate a power density of \SI{0.24}{\milli \watt / \meter \squared}. 

It may be noted that each \gls{esl} should have an energy buffer to store enough energy to support a single \gls{esl} update. Capacitors are most suited in this scenario. Suppose the harvester can boost the voltage with a \gls{mppt} boost converter to \SI{5}{\volt}, then the buffer capacitor will have to be larger than \SI{40}{\milli\farad} to store the \SI{0.5}{\joule} of energy. The self-discharge of the energy buffer could also be included in the estimations, especially when the energy buffer is recharged over a very long period of time. For the remainder of this paper we neglect the self-discharge of the energy buffer since this depends on the selected technology.

To calculate the necessary input RF power level from the required \gls{esl} DC energy, the harvester efficiency should be determined. Since there are many possible implementations of RF harvesters \cite{VanMulders2022} with variable efficiencies over the full input power range, the efficiency level is assumed to be constant for simplicity reasons. In the sequel, a harvester is assumed to have an efficiency of only \SI{30}{\percent}. Assuming that a constant net power of \SI{12}{\micro\watt} should be received and taking into account losses in the harvester conversion, there should continuously be at least \SI{-14}{\dBm} \gls{rf} input power, an input level that is above the input sensitivity of most harvesters. If the 600 \gls{esl} tags are powered sequentially, % and require screen update twice a day,% 
the charge time is limited to \SI{72}{\second} to store \SI{500}{\milli\joule} of energy in the buffer. This corresponds to requiring \SI{7}{\milli\watt} of power to be received during each time period. Assuming a harvester efficiency of \SI{30}{\percent}, the received RF power should be \SI{23}{\milli\watt} or approximately \SI{14}{\dBm}.

In the subsequent sections, we focus on the required transmit energy for charging the \gls{esl} buffers in array near field scenarios. In this feasibility study, we consider various charging options, as explained in \cref{subsection:options}. The results of this analysis are obtained through calculations aimed at quickly estimating the necessary transmit powers for the 351 antennas. A further fine-tuned analysis is conducted through a simulation framework that uses a geometry-based spherical wavefront channel model that includes the \gls{los} and represents \glspl{smc} at large planar surfaces by image sources. For this purpose, we have reformulated the Friis transmission equation in terms of power wave amplitudes, and thus the entries of the channel vector as \glspl{s-parameter}. Using \gls{vna} measurements, we have verified that our channel model is in close agreement with real-life measurements and allows us to simulate received signal powers in a physically correct manner~\cite{Deutschmann23ICC}.

\begin{table*}[t]
\centering
\caption{Estimated total required transmit power to energize the closest or furthest out of 600 \glspl{esl} for the SISO, MISO non-coherent and MISO coherent options. }%\jarne{de efficiency bij non coherent is gerelateerd aan één indivduele ESL, alhoewel verondesteld is dat meerder ESLs het vermogen zullen ontvangen. De efficientie zou dan opnieuw benaderend kunnnen berekend worden door rekening te houden met het aantal ESLs}} %\jarne{nakijken, ergens vermelden dat de coherent en non coherent waarden uit de tabel het totale vermogen is. om het individuel vermogen per transmitter te kennen moet je verminderen met 10log(351). Met deze informatie kunnen verdere hardware keuzes gemaakt worden, meer specifiek de PA technologie geselecteerd worde,. Merk op dat bij het siso systeem de individuel PA vermogens vele malen hoger liggen dan de coherent en non-coherent case.}}%Simplified version tentative}
\label{tab:receivedPower}
\setlength\tabcolsep{4pt}
\bgroup
\def\arraystretch{1.2}%
\begin{tabular}{@{\extracolsep{4pt}}llcccccclclclc@{}}
\hline
 & \multicolumn{1}{c}{} & \multicolumn{4}{c}{SISO} & \multicolumn{8}{c}{MISO} \\
 &  & \multicolumn{1}{l}{} & \multicolumn{1}{l}{} & \multicolumn{1}{l}{} & \multicolumn{1}{l}{} & \multicolumn{4}{c}{Non Coherent} & \multicolumn{4}{c}{Coherent} \\
 &  & \multicolumn{2}{c}{Calculated} & \multicolumn{2}{c}{Simulated} & \multicolumn{2}{c}{Calculated} & \multicolumn{2}{c}{Simulated} & \multicolumn{2}{c}{Calculated} & \multicolumn{2}{c}{Simulated} \\

Antenna & \begin{tabular}[c]{@{}l@{}}ESL-\\ Location\end{tabular}  & \multicolumn{1}{l}{\begin{tabular}[c]{@{}l@{}}$P_{tx,t}$\\ {[}dBm{]}\end{tabular}} & \multicolumn{1}{l}{$\eta$ {[}\%{]}}  & \multicolumn{1}{l}{\begin{tabular}[c]{@{}l@{}}$P_{tx,t}$\\ {[}dBm{]}\end{tabular}} & \multicolumn{1}{l}{$\eta$ {[}\%{]}}  & \multicolumn{1}{l}{\begin{tabular}[c]{@{}l@{}}$P_{tx,t}$\\ {[}dBm{]}\end{tabular}} & \multicolumn{1}{l}{$\eta$ {[}\%{]}}  & \multicolumn{1}{l}{\begin{tabular}[c]{@{}l@{}}$P_{tx,t}$\\ {[}dBm{]}\end{tabular}} & \multicolumn{1}{l}{$\eta$ {[}\%{]}}  & \multicolumn{1}{l}{\begin{tabular}[c]{@{}l@{}}$P_{tx,t}$\\ {[}dBm{]}\end{tabular}} & \multicolumn{1}{l}{$\eta$ {[}\%{]}}  & \multicolumn{1}{l}{\begin{tabular}[c]{@{}l@{}}$P_{tx,t}$\\ {[}dBm{]}\end{tabular}} & \multicolumn{1}{l}{$\eta$ {[}\%{]}} \\ \cline{3-6} \cline{7-10} \cline{11-14}
% Dipole & Closest & 49.9 & 0.025 & 45.7 & \SI{-31.74}{\dB} & 27.8 & 0.007 & 21.6 & \SI{-35.64}{\dB} & 28.4 & 3.7 & 29.8 & \SI{-15.81}{\dB} \\
% Dipole & Furthest & 60.9 & 0.0002 & 52.2 & \SI{-38.16}{\dB} & 34.8 & 0.001 & 31.0 & \SI{-45.01}{\dB} & 35.6 & 0.7 & 37.1 & \SI{-23.06}{\dB} \\
% Patch & Closest & 43.3 & 0.12 & 46.5 &  \SI{-32.50}{\dB} & 23.4 & 0.018 & 18.3 & \SI{-32.30}{\dB}  & 21 & 20.0 & 26.2 & \SI{-12.23}{\dB} \\
% Patch & Furthest & 42.9 & 0.13 & 57.6 & \SI{-43.64}{\dB} & 26.7 & 0.009 & 21.3 & \SI{-35.30}{\dB} & 23.8 & 10.5 & 29.6 &  \SI{-15.62}{\dB} \\ \hline
Dipole & Closest & 49.9 & 0.025 & 45.7 & 0.07 & {\color{lightgray} 27.8} & {\color{lightgray} 0.007 } & {\color{lightgray} 27.23} & {\color{lightgray} 0.008 } & 28.4 & 3.7 & 29.8 & 2.62 \\
Dipole & Furthest & 60.9 & 0.0002 & 52.2 & 0.02 & 34.8 & 0.001 & 34.48 & 0.001 & 35.6 & 0.7 & 37.1 & 0.49 \\
Patch & Closest & 46.8 & 0.05 & 46.5 &  0.06 & {\color{lightgray} 23.4 } & {\color{lightgray} 0.018 } & {\color{lightgray} 23.65 } & {\color{lightgray} 0.017} & 21 & 20.0 & 26.2 & 5.98 \\
Patch & Furthest & 47.2 & 0.05 & 57.6 & 0.004 & 26.7 & 0.008 & 27.04 & 0.008 & 23.8 & 10.5 & 29.6 & 2.74 \\ \hline
\end{tabular}
\egroup
% \vspace{4pt}

 \vspace{-12pt}
\end{table*}

\subsection{Coherent Massive Beamforming to Boost Efficiency}
\label{subsection:options}
A general claim for RF-based \gls{wpt} is the energy efficiency gain MISO systems have to the detriment of higher initial, installation and maintenance costs. The question is if this still stands for dense scenarios such as in the \glspl{esl} case and if there is a significant benefit when using coherent massive beamforming. Three provisioning options are shortly discussed here, two with \gls{siso} operation and two with \gls{miso}. These options require different levels of \gls{csi}~\cite{Lopez2019}, where a better knowledge of the \gls{miso} channel $\bm{h}\in\mathbb{C}^{L}$ allows to achieve higher performance.

\subsubsection{Option 1, Multiple Radio Elements with a Single Antenna} 
Radio elements are distributed above the aisle where sequentially each \gls{esl} receiver is powered solely by the closest radio element. A coarse-grained localisation system of the \glspl{esl} can provide the information to select for each receiver the corresponding transmitting radio element. In our further discussion, specifically two of the 600 \glspl{esl} are considered, the closest and furthest. We acknowledge that due to high path losses, a sequential \gls{siso} method is not feasible with a high number of \gls{en} devices. Still, the results from this method serve as a baseline for what is possible with single radio elements.
%In this \gls{siso} energy provision option, interference between multiple transmit signals is considered. \bert{What do you mean with considered? Is this implemented? How is this considered?} 
In addition, there is a trade-off between taking the closest transmit antenna or a better located transmit antenna due to more predominant losses related to the antenna radiation patterns. A post-processing step involving a sweep over all transmit antennas may therefore be required, wherein the path loss for each \gls{re} to the corresponding \gls{esl} could be estimated. 
%\jarne{We clarify that this \gls{siso} option is a baseline case with the assumption that only one \gls{esl} needs to be supported per day. A sequential process, as further considered in option 3, is not feasible with a \gls{siso} system due to the very high path loss. In the subsequent analysis, it is assumed that a constant received net power of \SI{-14}{\dBm} is required to update an \gls{esl} screen.}
Apart from choosing the closest located antenna, \gls{siso} operation does not assume knowledge of the channel and is hence regarded as a \textit{\gls{csi}-free} method. 

\subsubsection{Option 2, Non-coherent transmission by multiple distributed radio elements} A single radio element can have multiple antennas in this scenario. Several unsynchronized radio elements aim to deliver a quasi-uniform power density throughout the aisle to avoid blind spots, making sure that each \gls{esl} receives sufficient energy to perform the necessary updates. Interference of the \gls{rf} signals coming from the transmitters will occur randomly. \Cref{eq:prx_noncoherent} approximates the expected receive power for a non-coherent system $\mathbb{E} \big\{ P_{rx,nc} \big\}$ with $L$-number of antennas and a total transmit power of $P_{tx,t}$.
\setlength{\abovedisplayskip}{3pt}
\begin{equation}
    \mathbb{E} \big\{ P_{rx,nc} \big\} = % \approx 
    %G_\text{nc} 
    %\frac{P_\text{tx,t}}{L} 
    \sum_{l=1}^{L} P_{tx,l} 
    \, G_{tx,l}(\theta_l,\phi_l) \, G_{rx}(\theta_l,\phi_l) \, \left(\frac{\lambda}{4\pi d_l}\right)^2
    \label{eq:prx_noncoherent}
\end{equation}
We have shown that random beamforming (random channel amplitudes and phases), a \gls{csi}-free method, performs on average no more efficient than using an equivalent \gls{siso} system, and thus leverages no array gain~\cite{Deutschmann23ICC}. %\gilles{Just to make sure, you do get a received power gain when using MISO instead of SISO because you have more overall transmit power available. However, given that you also have a maximum power density rule, the MISO becomes indeed the SISO because the transmit power at all antennas needs to be lowered to respect this rule, i.e., PtxSISO / L  = PtxMISO. This also highly depends on what you consider the SISO case. Here, I have assumed, to have a fair comparison, that you average over all antennas in the SISO case (same tx antennas in the MISO but not on at the same time). Thus, this implies both (SISO and MISO) are the same of we have an overall receive power constraint (EIRP for example).} 
In this work we employ \textit{uniform} transmit powers $P_{tx,i}=P_{tx,j}=\frac{P_{tx,t}}{L} \, \forall \, i,j \in \{1 \dots L \}$, with $P_{tx,t}$ being the total transmit power of all $L$ antennas, which is likewise a \gls{csi}-free method that leverages no array gain.
This stands in contrast to non-coherent beamforming methods that rely on \textit{partial \gls{csi}} (e.g., channel magnitudes $|h_l|$) and leverage some array gain~\cite{REINDEERGeneral}. %~\cite[eq.\,(4.8)]{D2_3}. % We could remove this sentence/reference in the initial submission...
%However, the non-coherent transmission (assuming only random phases) inherently modeled in~\eqref{eq:prx_noncoherent} will leverage some \textit{noncoherent gain} $G_\text{nc}$ as a result of some antennas being closer to the \gls{en} device than others, as we have derived in ~\cite{REINDEERGeneral}. \gilles{REINDEERGeneral reference is not specific enough} \gilles{Mention that an additional gain can be achieved when the channel gain, or amplitudes are known if and only if, the transmit gains are limited by the power density regulations.} %\bert{This reference is still unpublished, but we would add the appendix in the final paper}.%Appendix~\ref{sec:appendix}. %or in~\cite[eq.\,(4.8)]{D2_3}. 
%The non-coherent \gls{miso} operation thus assumes knowledge of the channel magnitudes $|h_l|$ and is hence regarded as a method relying on \textit{partial \gls{csi}}.\newline
The latter part of~\eqref{eq:prx_noncoherent} represents the simulated path loss model based on the Friis transmission equation. 
$G_{tx,l}$ and $G_{rx}$ are respectively the transmit and receive antenna gain and $\theta_l$ and $\phi_l$ the angles related to the incident RF beam that differs for each transmit antenna location. $d_l$ represents the Euclidean distance between the transmit antenna and the \gls{en} device. We here neglect potential losses due to polarization mismatch. Referring back to the \gls{esl} case, all devices must receive a power level above the threshold of \SI{-14}{\dBm} to make the non-coherent solution achievable.
\subsubsection{Option 3: Coherent transmission by multiple distributed transmitters} In this scenario radio elements can generate power spots in the near field~\cite{Lopez2022} at the receiver location. %This means that, just like in option 1, location information on the \gls{esl} should be known. %In this option, a single radio element can again have multiple antennas and different radio elements 
In this option, all radio elements should be synchronized in time, phase, and frequency. This extra complexity can be overcome by reciprocity based calibration of the \gls{rf} front-ends \cite{REINDEERGeneral}. \Cref{eq:prx_coherent} represents the receive power estimation $P_{rx,c}$ with perfect constructive combination at the location of the \gls{en} device.
\begin{equation}
    P_{rx,c} = \left(\sum_{l=1}^{L} \sqrt{P_{tx,l} \, G_{tx,l}(\theta_l,\phi_l) \, G_{rx}(\theta_l,\phi_l) \, \left(\frac{\lambda}{4\pi d_l}\right)^2} \right)^2
    \label{eq:prx_coherent}
\end{equation}
The fully coherent \gls{miso} operation assumes \textit{perfect} knowledge of the channel vectors $\bm{h}$ in amplitude and phase and is hence regarded as a method demanding \textit{full \gls{csi}}.
Similar to the \gls{siso} case, each \gls{esl} is required to receive a power level of \SI{14}{\dBm} for \SI{72}{\second} to replenish the \gls{esl} buffer with energy. Subsequently, the power spot will be swept across all 600 \glspl{esl} locations.

\section{Results}\label{sec:results}
In the supermarket aisle feasibility study, two representative \gls{esl} locations are considered. The worst-located \gls{esl} is situated at a corner side, the best-located \gls{esl} can be found in a central location, both depicted respectively in blue and in red in Fig.\,\ref{fig:aislestore}. We determined the gain of the antennas based on the transmit antenna locations and the \gls{esl} placements. Two well-known antenna designs are considered: a quasi-omnidirectional dipole and a directional patch antenna. Their radiation patterns in combination with the changing angle of incidence for each transmit-receiver pair change the linear gain. Note that the coupling between the antennas, which may be considerable in the case of dipoles, is not considered.
% The results for the best and worst \gls{esl} in combination with each of the 351 transmit antennas is depicted in \bert{add figure.} \bert{Maybe we can add the figure in a later phase, where we can submit a 4 page paper? For now, it would take too much space.} \jarne{akkoord}
%Table\,\ref{tab:receivedPower} outlines three parameters for the two types of antennas at the two \gls{esl} locations: (i) the received power levels($P_{rx}$) when each antenna element sends out a \SI{0}{\dBm} signal. (ii) the total transmitted power ($P_{tx,t}$) to receive the \SI{-14}{\dBm} input power at the \gls{esl} in case of the \gls{siso} and non-coherent \gls{miso}, and \SI{14}{\dBm} for the coherent massive beamforming in case of 600 \glspl{esl}. The total transmitted power ($P_{tx,t}$) can be calculated with:
\Cref{tab:receivedPower} provides two parameters for the two types of antennas at the two \gls{esl} locations: (i) the total required transmit power ($P_{tx,t}$) to receive the \SI{14}{\dBm} input power at the \gls{esl} in case of the \gls{siso} and coherent \gls{miso} massive beamforming in case of 600 \glspl{esl}, and \SI{-14}{\dBm} for the non-coherent case. The relation between the total transmit power ($P_{tx,t}$) and the transmit power of each individual antenna ($P_{tx,l}$) is given by %\cref{eq:pwr_relation}.
\begin{equation}
P_{tx,t} = P_{tx,l} + 10 \log(L) \, ,
\label{eq:pwr_relation}
\end{equation}
where $L$ represents the number of antennas and $P_{tx,l}$ the transmit power per antenna. Through this formula, an estimate of the required power that the hardware must support can be derived, facilitating a more precise selection of the \gls{pa} \cite{VanMulders2022}. (ii) the efficiency or inverse path loss of the system, which is the ratio between the received power and the total transmitted power. The two parameters are calculated in 
idealised conditions, assuming \gls{los} and no multi-path reflections. %, and simulated based on data from measurements of real channels. 
The simulated data was derived by utilizing historical channel measurements, as previously stated in the \cref{subsection:deploy}. Furthermore, the influence of the shelves is taken into account by the simulation environment, in contrast to the calculated values where this was neglected. From \cref{tab:receivedPower}, the key takeaway is that the coherent \gls{miso} case has a higher efficiency whilst the total transmit power falls within the regulations~\cite{ETSI_EN_302_208}. In the non-coherent case, the expected uniform received power enforces the transmit power to be that of the furthest \gls{esl}. Therefore, the closest \gls{esl} is being greyed out. Comparing the \gls{miso} options with the \gls{siso} solution shows that increasing the number of simultaneously transmitting antennas reduces the total required transmit power drastically. Note that the efficiency in \cref{tab:receivedPower} is determined per \gls{esl}. By creating a quasi-uniform field throughout the aisle, the approximated total efficiency will be around $600$ times higher for this use case, and will end up higher compared to the \gls{siso} operation. Note that polarisation losses are not accounted for in this assessment, yet they could be easily added either by interpreting a specific deployment scenario or by simply assuming a \SI{3}{\dB} loss occurring when linearly polarized antenna(s) are used on one side of the link, and circularly polarized antenna(s) on the other side, as is commonly done in \gls{rfid}.

\begin{figure}[tbh]
    \centering
    \includegraphics[width=0.40\textwidth]{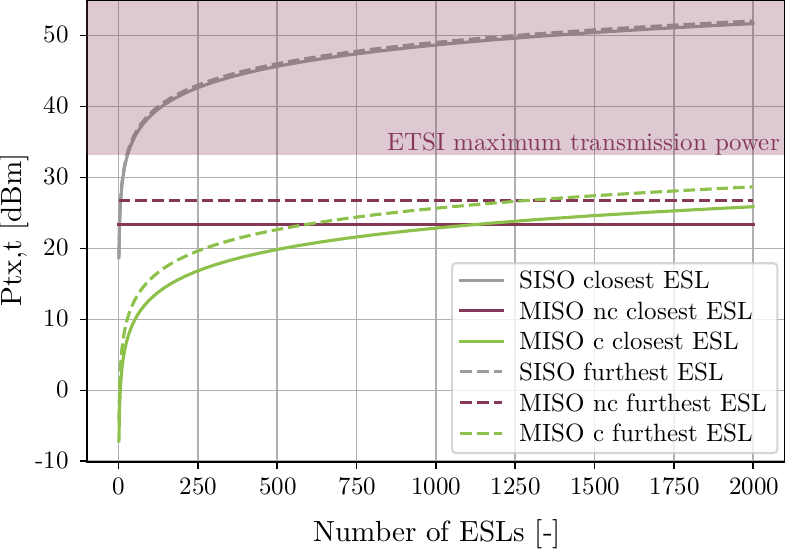}
    \vspace{-0.3cm}
    \caption{Evaluation of both SISO, non-coherent and coherent charging cases in relation to the number of \glspl{esl} within a aisle for the closest and furthest locations and with patch antennas.}
    \label{fig:transmitpowerNumberofESLs}
\end{figure}

Fig.\,\ref{fig:transmitpowerNumberofESLs} shows the relation between the required total antenna array transmit power and the number of \glspl{esl} for the three options when patch antennas are used. In the coherent \gls{miso} case, moving along the $x$-axis means that a higher number of \glspl{esl} should be powered in the same amount of time. To fill the energy buffers, a higher total transmit power is required. For the non-coherent \gls{miso} option, a constant, quasi-uniform field is already present, meaning that adding \glspl{esl} to the aisles does not influence the charge time and thus transmit power.  Depending on the amount of \glspl{esl}, the point where non-coherent \gls{miso} becomes more efficient than coherent massive beamforming lies at 1269 \glspl{esl}, worst case. Going from a \gls{siso} to a \gls{miso} scenario lowers the transmit power, with \SI{30.9}{\dBm} at the worst case scenario when patch antennas are used. The purple zone in Fig.\,\ref{fig:transmitpowerNumberofESLs} shows the point from where the maximum total transmit power violates the ETSI standard~\cite{ETSI_EN_302_208} in the \SI{868}{\mega\hertz} band. We want to stress again that the \gls{siso} option solely serves as a baseline and is not feasible in the supermarket aisle scenario, as it infringes these regulations already from a low number of \glspl{esl}. The required power increases rapidly due to high path losses, leading to unrealistic total transmit powers. For the plotted number of \glspl{esl}, both \gls{miso} options lie within this regulation zone. 
\section{Conclusion and future work}
\label{sec:conclusions}
In this study, a comparison between non-coherent quasi-uniform power density and coherent massive beamforming was conducted for a physically large antenna infrastructure in a practical supermarket aisle setup with 600 \glspl{esl}. % and non-coherent \gls{miso} was compared to the explained \gls{esl} use case was explored in a practical supermarket aisle setup. 
The feasibility of these methods was investigated by estimating the necessary transmit power levels of the $351$ transmit antennas, considering the required power at the \gls{esl} with two screen updates per day, the harvester efficiency, the antenna radiation patterns, and channel models. It was demonstrated that \gls{siso} systems do not hold up in these use cases due to the high path losses leading to extensive transmit power levels exceeding regulations. Conversely, the results predict sufficiently high gains for both \gls{miso} operations to support several hundreds of \glspl{esl} with the proposed transmit antenna configuration. As the number of \glspl{esl} increases, a non-coherent system may even become more efficient than the coherent one (cf.\,\cite{Lopez2019}), although in the latter beamsharing, which was neglected in current analysis, will likely be present in reality as neighboring \glspl{esl} receive a portion of the power from the same focal point. Accounting for beamsharing and/or multi-beam transmission in future work can further improve the accuracy of the estimates and the total efficiency, potentially making the coherent approach more efficient for a larger quantity of \glspl{esl}. The time, frequency, and phase synchronization required in this coherent approach should not be underestimated. Feasibility will be further validated in the real-world Techtile testbed \cite{REINDEERGeneral} %\bert{bron toevoegen?} 
located in Ghent, Belgium to apply the reciprocity-based beamforming approach and measure the non-coherent and coherent power levels. Additionally, the carrier frequency could be slightly increased to the \SI{917.5}{\mega\hertz} frequency band, where the transmission power could be increased to \SI{4}{\watt}.% double what is allowed in the \ band."

\renewcommand*{\bibfont}{\footnotesize}
\printbibliography
% \balance

% \bibitem{b1} G. Eason, B. Noble, and I. N. Sneddon, ``On certain integrals of Lipschitz-Hankel type involving products of Bessel functions,'' Phil. Trans. Roy. Soc. London, vol. A247, pp. 529--551, April 1955.
% \bibitem{b2} J. Clerk Maxwell, A Treatise on Electricity and Magnetism, 3rd ed., vol. 2. Oxford: Clarendon, 1892, pp.68--73.
% \bibitem{b3} I. S. Jacobs and C. P. Bean, ``Fine particles, thin films and exchange anisotropy,'' in Magnetism, vol. III, G. T. Rado and H. Suhl, Eds. New York: Academic, 1963, pp. 271--350.
% \bibitem{b4} K. Elissa, ``Title of paper if known,'' unpublished.
% \bibitem{b5} R. Nicole, ``Title of paper with only first word capitalized,'' J. Name Stand. Abbrev., in press.
% \bibitem{b6} Y. Yorozu, M. Hirano, K. Oka, and Y. Tagawa, ``Electron spectroscopy studies on magneto-optical media and plastic substrate interface,'' IEEE Transl. J. Magn. Japan, vol. 2, pp. 740--741, August 1987 [Digests 9th Annual Conf. Magnetics Japan, p. 301, 1982].
% \bibitem{b7} M. Young, The Technical Writer's Handbook. Mill Valley, CA: University Science, 1989.
% \end{thebibliography}
% \vspace{12pt}
% \color{red}
%IEEE conference templates contain guidance text for composing and formatting conference papers. Please ensure that all template text is removed from your conference paper prior to submission to the conference. Failure to remove the template text from your paper may result in your paper not being published.

\end{document}

%% file: abbr.tex
%
% Add new acronyms in alphabetical order
%
\newacronym{ai}{AI}{Artificial Intelligence}
\newacronym{aoa}{AOA}{angle-of-arrival}
\newacronym{aod}{AOD}{angle-of-departure}
\newacronym{ap}{AP}{access point}
\newacronym{asic}{ASIC}{application specific integrated circuit}
\newacronym{awgn}{AWGN}{additive white Gaussian noise}
\newacronym{ce}{CE}{channel estimation}
\newacronym{cw}{CW}{continuous wave}
\newacronym{ced}{CED}{cumulative energy density}
\newacronym{csi}{CSI}{channel state information}
%\newacronym[plural=csps,firstplural=contact service points (CSPs)]{csp}{CSP}{contact service point}
\newacronym{csp}{CSP}{contact service point}
\newacronym{crlb}{CRLB}{Cram\'er-Rao lower bound}
\newacronym{dc}{DC}{direct current}
\newacronym{dlt}{DLT}{Distributed Ledger Technology}
\newacronym{dp}{DP}{differencial privacy}
\newacronym{dpd}{DPD}{digital pre-distortion}
\newacronym{ecc}{ECC}{elliptic curve cryptography}
\newacronym{ecdlp}{ECDLP}{elliptic curve discrete logarithm problem}
\newacronym{ecsp}{ECSP}{edge computing service point}
\newacronym{eh}{EH}{energy harvesting}
\newacronym{eirp}{EIRP}{equivalent isotropically radiated power}
\newacronym{em}{EM}{electromagnetic}
\newacronym{en}{EN}{energy neutral}
\newacronym{e2e}{E2E}{End-to-End}
\newacronym{end}{E2E}{energy neutral device}
%\newacronym{epu}{EPU}{edge processing unit} %Not used. Changed to ECSP.
\newacronym{fa}{FA}{federation anchor}
\newacronym{fhss}{FHSS}{frequency hopping spread spectrum}
\newacronym{fl}{FL}{federated learning}
\newacronym{gdpr}{GDPR}{general data protection regulation}
\newacronym{ic}{IC}{integrated circuit}
\newacronym{ioe}{IoE}{Internet of Everything}
\newacronym{iot}{IoT}{Internet of Things}
\newacronym{id}{ID}{information decoding}
\newacronym{kpi}{KPI}{key performance indicator}
\newacronym{lca}{LCA}{life cycle assessment}
\newacronym{lis}{LIS}{large intelligent surface}
\newacronym{liion}{Li-ion}{lithium-ion}
\newacronym{los}{LoS}{line-of-sight}
\newacronym{lmo}{LMO}{lithium ion manganese oxide}
\newacronym{mc}{MC}{Monte Carlo}
\newacronym{mf}{MF}{matched filter}
\newacronym{ml}{ML}{machine learning}
\newacronym{mmse}{MMSE}{minimum mean square error}
\newacronym{mmwave}{mmWave}{millimeter wave}
\newacronym{mimo}{MIMO}{multiple-input multiple-output}
\newacronym{mmimo}{mMIMO}{massive \acrshort{mimo}}
\newacronym{miso}{MISO}{multiple-input single-output}
\newacronym{mpc}{MPC}{multipath component}
\newacronym{mrc}{MRC}{maximum ratio combining}
\newacronym{mrt}{MRT}{maximum ratio transmission}
\newacronym{ofdm}{OFDM}{orthogonal frequency-division multiplexing}
\newacronym{ota}{OTA}{over-the-air}
\newacronym{nb}{NB}{narrowband}
\newacronym{nr}{NR}{New Radio}
\newacronym{pa}{PA}{power amplifier}
\newacronym{pdf}{PDF}{probability density function}
\newacronym{pdp}{PDP}{power delay profile}
\newacronym{per}{PER}{packet error rate}
\newacronym{pet}{PET}{privacy enhancing technology}
\newacronym{pki}{PKI}{public key infrastucture}
\newacronym{pl}{PL}{path loss}
\newacronym{pc}{PC}{pilot count}
\newacronym{re}{RE}{radio element}
\newacronym{rf}{RF}{radio frequency}
\newacronym{rfid}{RFID}{radio frequency identification}
\newacronym{ris}{RIS}{reflective intelligent surface}%reconfigurable?
\newacronym{rms}{RMS}{root mean square}
\newacronym{rss}{RSS}{received signal strength}
\newacronym{rssi}{RSSI}{received signal strength indicator}
\newacronym{rv}{RV}{random variable}
\newacronym{rw}{RW}{RadioWeaves}
\newacronym{sa}{SA}{synchronization anchor}
\newacronym{sdg}{SDG}{Sustainable Development Goal}
\newacronym{sdn}{SDN}{software-defined network}
\newacronym{sinr}{SINR}{signal-to-interference-plus-noise ratio}
\newacronym{siso}{SISO}{single-input single-output}
\newacronym{slam}{SLAM}{simultaneous localization and mapping}
\newacronym{smc}{SMC}{specular multipath component}
\newacronym{snr}{SNR}{signal-to-noise ratio}
\newacronym{snir}{SNIR}{signal-to-interference-plus-noise ratio}
\newacronym{svd}{SVD}{singular value decomposition}
\newacronym{s-parameter}{S-parameter}{scattering parameter}
%\newacronym{sppu}{SPPU}{service point processing unit}
%\newacronym{sp}{SP}{service point}
\newacronym{tdd}{TDD}{time division duplexing}
\newacronym{tdoa}{TDOA}{time-difference-of-arrival}
\newacronym{toa}{TOA}{time-of-arrival}
\newacronym{ue}{UE}{user equipment}
\newacronym{uhf}{UHF}{ultra high frequency}
\newacronym{ula}{ULA}{uniform linear array}
\newacronym{upa}{UPA}{uniform planar array}
\newacronym{ura}{URA}{uniform rectangular array}
\newacronym{uwb}{UWB}{ultrawideband}
\newacronym{vna}{VNA}{vector network analyzer}
\newacronym{wb}{WB}{wideband}
\newacronym{wpt}{WPT}{wireless power transfer}
\newacronym{zf}{ZF}{zero forcing}

\newacronym{lpt}{LPT}{laser power transfer}
\newacronym{rfpt}{RFPT}{radio frequency power transfer}
\newacronym{ipt}{IPT}{inductive power transfer}
\newacronym{cpt}{CPT}{capacitive power transfer}
\newacronym{apt}{APT}{acoustic power transfer}
\newacronym{cfo}{CFO}{carrier frequency offset}
\newacronym{lo}{LO}{local oscillator}
\newacronym{if}{IF}{intermediate frequency}
\newacronym{duc}{DUC}{digital up conversion}
\newacronym{ddc}{DDC}{digital down conversion}
\newacronym{dsp}{DSP}{digital signal processing}
\newacronym{vco}{VCO}{voltage-controlled oscillator}
\newacronym{pll}{PLL}{phase-locked loop}
\newacronym{lna}{LNA}{low noise amplifier}
\newacronym{dac}{DAC}{digital-to-analog converter}
\newacronym{adc}{ADC}{analog-to-digital converter}
\newacronym{de}{DE}{drain efficiency}
\newacronym{pae}{PAE}{power-added efficiency}
\newacronym{sar}{SAR}{specific absorption rate}
\newacronym{mppt}{MPPT}{maximum power point tracking}
\newacronym{isi}{ISI}{intersymbol interference}
\newacronym{pps}{PPS}{pulse per second}
\newacronym{icnirp}{ICNIRP}{International Commission on Non-Ionizing Radiation Protection}
\newacronym{esl}{ESL}{electronic shelf label}
\newacronym{edlc}{EDLC}{electrostatic double-layer capacitors}
\newacronym{papr}{PAPR}{peak-to-average power ratio}
\newacronym{tcxo}{TCXO}{temperature compensated crystal oscillator}
\newacronym{rcs}{RCS}{radar cross section}
\newacronym{ook}{OOK}{on-off keying}
\newacronym{ask}{ASK}{amplitude shift keying}
\newacronym{sdma}{SDMA}{spatial-division multiple access}
\newacronym{fdma}{FDMA}{frequency-division multiple access}
\newacronym{cdma}{CDMA}{code division-multiple access}
\newacronym{tdma}{TDMA}{time division-multiple access}
\newacronym{pn}{PN}{pseudo-noise}
\newacronym{rtf}{RTF}{reader talks first}
\newacronym{ttf}{TTF}{tag talks first}
\newacronym{ber}{BER}{bit error rate}
\newacronym{rir}{RIR}{reverse-link interrogation range}
\newacronym{oma}{OMA}{orthogonal multiple access}
\newacronym{noma}{NOMA}{nonorthogonal multiple access}
\newacronym{pe}{PE}{processing element}
\newacronym{dsb}{DSB}{double-sideband}
\newacronym{ssb}{SSB}{single-sideband}
\newacronym{pr}{PR}{phase reversal}
\newacronym{lpf}{LPF}{low-pass filter}
\newacronym{mcu}{MCU}{microcontroller}